\documentclass[journal]{IEEEtran}

\usepackage{xcolor,soul,framed} 
\usepackage{threeparttable}
\colorlet{shadecolor}{yellow}
\usepackage[pdftex]{graphicx}
\graphicspath{{../pdf/}{../jpeg/}}
\DeclareGraphicsExtensions{.pdf,.jpeg,.png}
\usepackage{caption}
\usepackage{booktabs}
\usepackage{makecell}
\usepackage{float}
\usepackage{graphicx}                                          \usepackage{stfloats}
\usepackage{seqsplit}

\usepackage[cmex10]{amsmath}
\usepackage{array}
\usepackage{mdwmath}
\usepackage{mdwtab}
\usepackage{eqparbox}
\usepackage{url}
\usepackage{amssymb} 
\usepackage{utfsym} 
\usepackage{microtype}
\usepackage{placeins}

\begin{document}
\begin{sloppypar}

\bstctlcite{IEEEexample:BSTcontrol}
  \title{Pathology-genomic fusion via biologically informed cross-modality graph learning for survival analysis}
  \author{Zeyu Zhang, Yuanshen Zhao, Jingxian Duan, Yaou Liu, Hairong Zheng, Dong Liang \\ Zhenyu Zhang and Zhi-Cheng Li
  
  \thanks{This work was funded by the National Natural Science Foundation of China U20A2017; the Key-Area Research and Development Program of Guangdong Province, China (No. 2021B0101420006); National Natural Science Foundation of China (No. 62201557, U20A20171, 12126608); R\&D project of Pazhou Lab (Huangpu) under Grant 2023K0603; Guangdong Basic and Applied Basic Research Foundation (No. 2021A1515110585), Shenzhen Medical Research Fund(A2303008)}

  \thanks{Corresponding Author: Zhi-Cheng Li: zc.li@siat.ac.cn; Zhenyu Zhang: fcczhangzy1@zzu.edu.cn}%
    
  \thanks{Zeyu Zhang, Yuanshen Zhao, Jingxian Duan, Hairong Zheng, Dong Liang and Zhi-Cheng Li are with Shenzhen Institute of Advanced Technology, Chinese Academy of Sciences, Shenzhen, 518055 China.}
  
  \thanks{Hairong Zheng, Dong Liang and Zhi-Cheng Li are also with the Key Laboratory of Biomedical Imaging Science and System, Chinese Academy of Sciences, Shenzhen, China}
      
  \thanks{Yaou Liu is with Beijing Tiantan Hospital, Capital Medical University, Beijing, 100162, China.}
  
  \thanks{Zhenyu Zhang is with the First Affiliated Hospital of Zhengzhou University, Zhengzhou, 480082 Henan, China.}

  }

\maketitle

\begin{abstract}
The diagnosis and prognosis of cancer are typically based on multi-modal clinical data, including histology images and genomic data, due to the complex pathogenesis and high heterogeneity. Despite the advancements in digital pathology and high-throughput genome sequencing, establishing effective multi-modal fusion models for survival prediction and revealing the potential association between histopathology and transcriptomics remains challenging. In this paper, we propose Pathology-Genome Heterogeneous Graph (PGHG) that integrates whole slide images (WSI) and bulk RNA-Seq expression data with heterogeneous graph neural network for cancer survival analysis. The PGHG consists of biological knowledge-guided representation learning network and pathology-genome heterogeneous graph. The representation learning network utilizes the biological prior knowledge of intra-modal and inter-modal data associations to guide the feature extraction. The node features of each modality are updated through attention-based graph learning strategy. Unimodal features and bi-modal fused features are extracted via attention pooling module and then used for survival prediction. We evaluate the model on low-grade gliomas, glioblastoma, and kidney renal papillary cell carcinoma datasets from the Cancer Genome Atlas (TCGA) and the First Affiliated Hospital of Zhengzhou University (FAHZU). For demonstrating the model interpretability, we also visualize the attention heatmap of pathological images and utilize integrated gradient algorithm to identify important tissue structure, biological pathways and key genes. 
\end{abstract}


%
\IEEEpeerreviewmaketitle



\section{Introduction}

\IEEEPARstart{C}{ancer} diagnosis and prognosis usually involve heterogeneous clinical data including histology images and genomic data owing to the complex pathogenesis and high heterogeneity \cite{beck2015open}. Despite the enormous advancement of digital pathology and high-throughput genome sequencing technology, it remains challenging to establish an effective multi-modal fusion model and identify the correlations between histopathology and transcriptomics.
Whole slide images (WSI) contain rich information about tissue structure (e.g., fibrous tissue, glandular tissue or blood vessels) and cell type (e.g., cancer cells, stromal cells or immune cells) which can illustrate the malignant degree and development stage of tumor. Consequently, the pathologists usually consider the pathological characteristics as an important criteria of cancer diagnosis and grading \cite{courtiol2019deep}. However, the subjective pathological descriptions fail to capture survival-associated features due to high heterogeneity of WSI and multifactorial influence of prognosis. Transcriptomics provides complementary prognostic information \cite{jaume2023modeling} including quantitative measurement of RNA expression levels which can reveal cancer pathogenesis and progression from a microscopic perspective. Meanwhile, according to the central dogma of molecular biology, gene expression can affect the protein translation, consequently influencing the cells growth and proliferation, which can be reflected as pathological features in WSI, thus gene expression patterns have the potential correlation with histopathological characteristics. The major challenge posed by the integration of both modalities for survival prediction is how to establish intra-modal and inter-modal correlations that are consistent with prior biological knowledge to achieve reasonable and effective fusion of genomic data and histology images.
Currently, multi-modal fusion strategies can be divided into early, late, and intermediate fusion. Early fusion integrates feature of different modalities through vector concatenation, bilinear pooling or element-wise sum before input the fused feature into a unified model \cite{le2017automated}, which requires feature alignment across different modalities. However, different modalities are usually complementary and intersectional \cite{wang2023shared}, especially genomic data and WSI that have huge data heterogeneity gap and vast spatial scales discrepancy \cite{chen2020pathomic,chen2022pan}. Late fusion allows for the specific feature extraction networks for each modality and then aggregates the output of each network \cite{shao2019integrative}. Although promising performance has been reported by numerous researches, the late fusion method usually ignores the correlation among different modalities due to the separated representation learning networks. Compared with early fusion and late fusion, the intermediate fusion integrates the representation learning and multi-modal fusion network, which means inter-modal information interaction in fusion layer can influent feature extraction of each modality in representation learning network under the multimodal context \cite{havaei2016hemis,lv2021pg,vale2021long}. Moreover, intermediate fusion enables the model to explore complex correlation across multi-modal data, which has substantial potential in improving the model interpretability. Cross-guided fusion is a frequently-used  intermediate fusion strategy which allow the guidance of feature extraction among different modalities. For instance, recent work has incorporated transformer-based architecture to multi-modal fusion task and model complex interaction across each modality via guided-attention mechanism \cite{jaume2023modeling,li2023survival}. However, transformer-based architecture calculate the co-attention matrix simply through matrix multiplication which ignore the intrinsic associations of each modality e.g., spatial information of WSI and biological correlation of biological pathways.
In this paper, we propose pathology-genome heterogeneous graph (PGHG) consisting of biological knowledge guided representation learning network and pathology-genome heterogeneous graph to integrate WSI and bulk RNA-Seq expression data for survival analysis. Specifically, pathway enrichment analysis is performed to obtain biological pathways which have been verified that involved in different physiological and metabolic processes in cells and have specific biological meanings, then we represent the pathways as nodes in the genome subgraph, and the topology relationship among each pathway node is established according to the common RNA number. Besides, whole slide image is divided into non-overlapping patches which are represented as nodes in pathology subgraph and the edges are built based on the adjacent relation of each patch. In the representation learning network, inspired by the graph auto-encoder that learns the node representations with structure information by reconstruct adjacent matrix of pathway subgraph, we further conduct RNA-Seq expression data reconstruction to make sure the characteristics of RNA expression are also preserved. Besides, given the central dogma of molecular biology and the potential associations between genomic data and histology image,we propose to utilize biological priori information to guide representation learning network to extract genome-relevant pathological feature and alleviate the interference of redundant features. Specifically, gene set variation analysis (GSVA) is applied to RNA-Seq data and the GSVA scores of each pathway are used to supervise the pathological feature extraction. Lastly, to reduce the large data heterogeneity gap and explicitly model the correlation between histology image and RNA-Seq expression data, we extract the global representation of each modality with attention pooling module and minimize the Euclidean distance of the global embeddings in the feature space. In the pathology-genome fusion heterogeneous graph, we construct the edges between pathology subgraph and genomic subgraph through fully connecting each heterogeneous node pairs, considering that bulk RNA-Seq expression data represents the comprehensive sequencing outcomes of cells within the entire sampled tissue. The heterogeneous graph learning is similar to graph attention networks and each node aggregates information from the intra-modal and inter-modal neighbor nodes consecutively, then we obtain the intra-modal feature and inter-modal feature of node in each subgraph which will be used in subsequent survival prediction.

Our contributions are summarized as follows:
(1) The representation learning module utilizes biological prior  knowledge to guide feature extraction of histology image and genomic data and align the feature embeddings from each modality which can decrease the heterogeneous gap of multi-modality.
(2) The heterogeneous graph construction is in accordance with biological prior knowledge and can model the correlation of multi-modal data through graph attention-based architecture which specifies the associations between biological pathways and histology image patches, presenting a intra-modal and inter-modal insight of model interpretability. 
(3) The heterogeneous graph neural network provides a novel perspective of biological guided intermediate fusion strategy which allows for the expansion of arbitrary number of data modalities and other graph construction strategy conforming to the specific characteristics of each modality.

\section{Related Work}

\subsection{Survival Analysis with Genomic features}Recently, the advancement in sequencing technologies greatly deepens the understanding of molecular biology, several genomic-based survival models utilizing molecular biomarkers and genomic expression data have been proposed. For instance, Shuguang Zuo et al. \cite{zuo2019rna} developed a COX regression model with six selected gene signatures for survival analysis of patients with glioblastoma and evaluated the relevant contribution of each gene signature for the survival prediction. However this strategy can only utilize limited number of selected signatures and failed to handle high-dimensional genomic profiles. Subsequently, recent works have incorporated deep neural network into genetic-based survival analysis. Safoora Yousefi et al. \cite{yousefi2017predicting} proposed a feedforward network driven by a Cox survival model with
mRNA, gene mutations, copy number variations (CNV), and
protein expression features as model input, besides the model hyperparameters were searched based on the Bayesian optimization. Zhi Huang et al. \cite{huang2019salmon} proposed to conduct gene co-expression network analysis and then used the identified co-expression modules and other cancer biomarkers for survival analysis, thus simplified high-dimensional gene expression data, provided a promising solution for improving the model interpretability and exploring the potential biological functions. However, these strategies fail to incorporate with other data modalities which may involve more comprehensive tumor-relevant prognostic information.

\subsection {Survival Analysis with histology image}
The combination of digital pathology and artificial intelligence has greatly accelerated the development of computational pathology, deep learning based diagnostic whole slide images analysis has been widely used in numerous clinical task such as cancer diagnosis, prognosis, therapeutic response predictions, etc. However, the training and deployment of conventional prognostic models remain challenging due to the gigapixels of WSI. Currently, the mainstream of survival analysis using WSIs is patch-based multiple instance learning. Gang Xu et al. \cite{xu2019camel} proposed a weakly supervised learning framework that leveraged image-level label to supervise the generation of instance-level labels, which were further assigned to the corresponding pixels and thus achieved pathological image segmentation. Besides, Richard J. Chen et al. \cite{chen2022pan} incorporate Attention MIL (AMIL) \cite{ilse2018attention} into cancer survival analysis with WSIs, which can generate the attention scores for the importance quantification of each image patch and identify the pathological characteristics in the high-attention regions. Later, they further proposed Hierarchical Image Pyramid Transformer (HIPT), a transformer-based hierarchical self-supervised framework for the representation learning of histological image, the model was pretrained with 10678 gigapixel WSIs from 33 cancer types and achieved remarkable performance in cancer subtyping and survival prediction. Despite the promising improvement MIL has achieved, treating each image patch as a separated instance neglects the morphological feature correlations between cell identities and pathological tissues, which may contain enormous prognostic information. Thus graph-based networks \cite{chen2021whole,shao2023characterizing} were proposed to address this problem, specifically, each histology image patch was regarded as a node in the graph and the edge was constructed based on the adjacency relation, thus the topological organization in the tumor microenvironment was considered in the graph learning.

\subsection {Survival Analysis with Multi-modal Learning}
With the emergence of clinical data in various modalities, the deep learning-based multi-modal fusion algorithm has become a promising strategy that seeks to aggregate multiple data modalities, improve prediction precision and explore multi-modal biomarkers. Richard J. Chen et al. \cite{chen2020pathomic} proposed to aggregate the histologic deep features, cell morphological characteristics and genomic features through Kronecker Product and gating-based attention mechanism. Subsequently, they also employed a simple late fusion mechanism: combining histology feature with genomic data through vector concatenation at the last layer and then evaluated the effectiveness of the strategy on 6,592 H\&E diagnostic WSIs of 14 cancer types from TCGA \cite{chen2022pan}. Besides, Transformer-based multi-modal fusion approaches have also shown great potential. There has been numerous studies on survival prediction using cross-attention module to model the correlation between different modalities \cite{li2023survival,chen2021multimodal,chen2021multimodal}, for instance, using a query generated from the genomic embedding to guide the histological feature extraction and thus generate genomic-guided histological image embeddings. However these approaches fail to make full use of potential biological prior knowledge and merely treat each data modalities as structured feature embedding, thus the intra-modality and inter-modality information are not taken into consideration.

\section{Methods}

The main framework of our proposed model is illustrated as Fig 1, which consists of biological knowledge guided representation learning module and pathology-genome heterogeneous graph module. In the representation learning module, we extract pathological and genomic feature embeddings and leverage the biological prior knowledge to capture potentially correlated information between pathological and genomic modalities. Specifically, we first divide WSI into non-overlapping patches after the image background
removal. Besides, pathway enrichment is conducted to identify statistically enriched biological pathways. Then we propose to leverage known biological information of intra-modality and inter-modality associations to guide the feature extraction of histological image patches and biological pathways, which are denoted as nodes in pathological subgraph and genomic subgraph respectively. For the reasonable and effective integration of pathological images and genomic data, we construct the pathological subgraph and genomic subgraph based on spatial location and common gene number respectively, pathological patch nodes are fully interconnected with pathway nodes to sufficiently model the intricate inter-modality interactions. In the heterogeneous graph module, the node embeddings are updated through attention-based graph learning \cite{velivckovic2017graph} and attention pooling module are utilized to extract the unimodal global feature and bi-modal fused global feature from each subgraph. Then the concatenated global feature is input into a fully connected layer to make the final survival prediction. 

\renewcommand\arraystretch{1.5} 
\begin{figure*}[!ht] 
\includegraphics[width=7.0in]{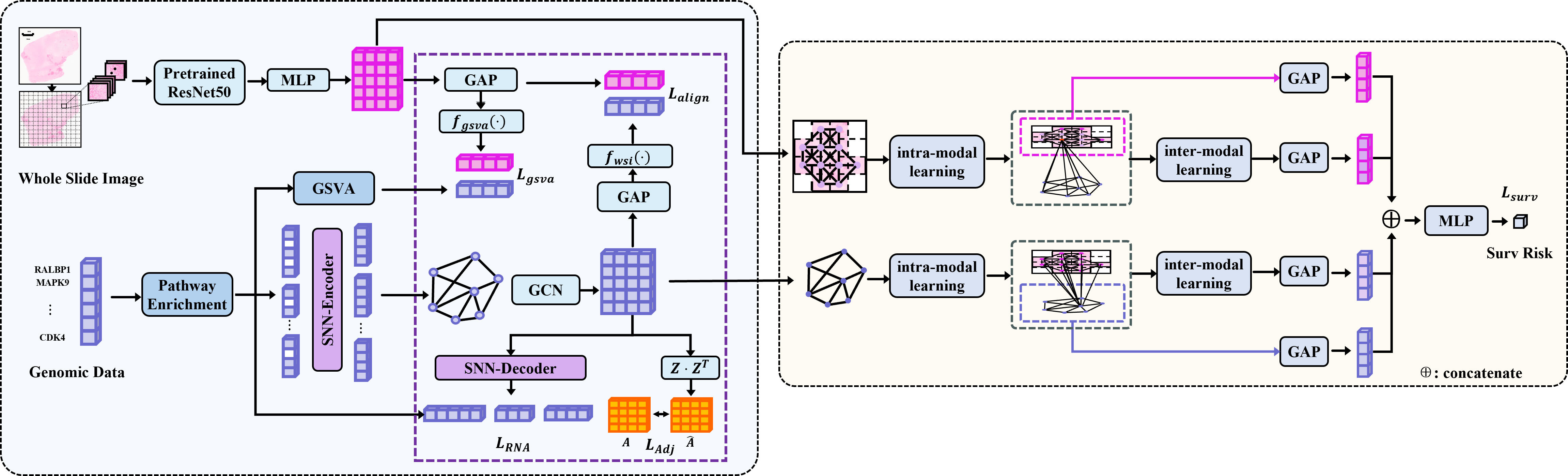}
\caption{Schematic illustration for the proposed method: pathology-genome heterogeneous graph. The blue box contains the biological knowledge guided representation learning network. The yellow box contains the pathology-genome heterogeneous graph.}
\label{amplifier}
\end{figure*}


\subsection {Pathological feature extraction and subgraph construction}
Computational pathology has achieved remarkable progress in multiple clinical task e.g. cancer detection \cite{liu2017detecting}, diagnosis \cite{coudray2018classification} and prognosis \cite{courtiol2019deep}. In recent years, an increasing number of researches adopted multiple instance learning (MIL) to gigapixel whole slide images analysis through representing each separated patch as an independent instance and integrating all the instances via global feature aggregation. Despite promising results have been reported, conventional MIL approaches may ignore the location information across patches and the interactions between cells and tissues remain indiscernible. Inspired by Patch-GCN \cite{chen2021whole}, we construct the pathology subgraph $G_p=\left(F_p,A_p\right)$ with pathological image, each patch is denoted as a node $f_p^i$ and connected to the 8 hop proximal neighbor patches $N\left(i\right)$. The adjacent matrix of pathology subgraph is denoted as $A_p$. Specifically, we divide each WSI into 256×256 patches at 20× magnification (0.5 $\mu$m/pixel) and extract feature embeddings which is denoted as the nodes feature $F_p=\left[f_p^0,f_p^1,\ldots,f_p^m\right]$ using ResNet50 pretrained on ImageNet. The adjacency matrix $A_p$ is defined as following formula:

\begin{equation*}
A_p [i,j]=
\begin{cases}
    1, & j\in N(i)\\
    0, & otherwise
\end{cases}
\end{equation*}

\subsection {Biological pathway feature extraction and subgraph construction}

Gene expression profiles are highly associated with cancer progression and may have the potential to provide fundamental explanation of cancer pathogenesis. Traditional techniques of transcriptome data analysis are usually based on statistical tests or machine learning which focus on differential gene expression analysis \cite{braun2008identifying} and identify the key genes, tumor-specific molecular mechanisms and candidate targets for drug therapy, etc. However, recent studies have indicated that the biological pathways are more conducive to revealing the pathogenesis of cancer \cite{liu2012identifying,zhang2017network,kim2015knowledge}. Thus we choose the biological pathway as the basic unit of genomic subgraph. Firstly, we identify the top 20000 genes with the highest variance from external genomic dataset, which are not involved in model training and validation, then we conduct pathway enrichment to high variance genes using biological pathway database KEGG and REACTOM. Pathways with statistical significance (adj P value $<$ 0.05) are reserved for the genomic subgraph construction. Subsequently, a logarithmic transformation for the gene expression values of each pathway is performed to reduce the impact of extreme values. The genomic subgraph $G_g=\left(F_g,A_g\right)$ is constructed in the following steps: (1) denote the biological pathway as the node and the sequencing data of genes contained in the each pathway are represented as the genomic node feature $F_g$; (2) we propose to quantify the correlation of biological function between pathways with the ratio of common genes number and the total genes number of both pathways. Each pathway node $f_g^i$ is connected to the top-10 pathway nodes that have the highest relevance score $s_{ij}\ \left(j\in\left[0,1,\ldots n\right]\right)$. The formulas are as follows:
\begin{equation*}
s_{ij} = \frac{num(f_g^i \cap f_g^j)}{num(f_g^i)+num(f_g^j)}
\end{equation*}
\begin{equation*}
A_g [i,j]=
\begin{cases}
    s_{ij}, & \ if\ s_{ij}\in top10\left(s_{ik},k\in \{1,2,\ldots,n\}\right)\\
    0, & otherwise
\end{cases}
\end{equation*}

\subsection {Biological prior knowledge guided representation learning}
\noindent \textbf{Genomic knowledge-guided pathway feature extraction}
In previous work, differential gene expression analysis of normal tissue and tumor tissue are usually conducted to identify the key genes which may relate to the molecular mechanism of cancer pathogenesis. However, the identified key genes may exhibit significant bias due to variations in statistical methods. Moreover, the functional dependency between genes or pathways which has been proved that have the potential to reveal the cancer pathogenesis remains unexploited \cite{braun2008identifying,liu2012identifying,curtis2005pathways}. Similar to the graph autoencoder \cite{kipf2016variational}, we conduct the reconstruction of pathway subgraph adjacent matrix and gene expression value to ensure the pathways correlation information and gene expression patterns can be preserved in the representation learning. Specifically, after the above-mentioned graph construction, we obtain the pathway subgraph represented as $G_g=\left(F_g,A_g\right)$. It's notable that the initial feature embedding of pathway node $f_g^i\ \left(i\in\left\{1,2,\ldots n\right\}\right)$ is composed of different number of genes and thus the dimension of pathway node features is not unified. Therefore, we adopt SNN encoder networks which consist of fully connected layer, exponential linear units (ELU) and Alpha Dropout to each pathway node to convert the features into dimension-unified embeddings $f_{gu}^i$. After the graph convolution operation, the pathway embeddings $Z$ are used to reconstruct the adjacent matrix $A_g$ via inner product. ${\widetilde{A}}_g=D^{-1/2}A_gD^{-1/2}$ represents the symmetrically normalized adjacency matrix and D is the degree matrix. The graph convolution and adjacent matrix reconstruction formulas are as follows:

\begin{equation*}
Z=ReLU({\widetilde{A}}_gReLU({\widetilde{A}}_gF_{gu}W_1)W_2)
\end{equation*}
\begin{equation*}
A_g^\prime=sigmoid(ZZ^T)
\end{equation*}
\begin{equation*}
    \begin{split}
    L_{Adj} = & -\frac{1}{N \times N} \Bigg(\sum_{i=0}^{N} \sum_{j=0}^{N} a_{ij} \log{(a_{ij}^{'})} \\
    & + (1 - a_{ij}) \log{(1 - a_{ij}^{'})}\Bigg)
    \end{split}
\end{equation*}

where the $a_{ij}$ represents the element in pathway subgraph adjacent matrix and the $a_{ij}'$ is the element of reconstructed adjacent matrix. $L_{Adj}$ is the cross-entropy loss of the graph reconstruction.\\
\indent Meanwhile, recent research indicates that the topological information of graph should not be over-emphasized while feature reconstruction with corruption is necessary to improve model robustness \cite{hou2022graphmae}, especially given the unstability of gene sequencing technique due to diverse sequencing depth, coverage and sample batch effect etc. Thus we additionally reconstruct the gene expression values $F_g$ to ensure the preservation of gene expression pattern in representation learning. Concretely, we input pathway node feature $f_{gu}^i$ into the decoder networks which have similar network structure with encoder network to convert the dimension-unified pathway embedding into the reconstructed pathway node feature $f_{gr}^i$ that has the identical dimension with $f_g^i$. To improve the generalization and robustness of model to unstable genomic data, we randomly mask 20$\%$ of the gene expression values within each pathway nodes in each epoch during the training process.\\
\indent For the pathway node $f_g^i$, $M$ denotes the gene index set that is randomly chosen to be masked, the masked pathway node $f_{gm}^i$ is defined as:
\begin{equation*}
f_{gm}^i[q] =
\begin{cases}
    f_g^{i}[q] & , q \in M \\
    0 & , q \notin M 
\end{cases}
\quad q \in [0, 1, \ldots, n]
\end{equation*}
$F_{recon}^i(\cdot)$ denote the decoder network of pathway node $i$. The reconstruction of gene expression values $F_g$ are defined as follows:\\
\begin{equation*}
f_{gr}^i=F_{recon}^i(f_{gu}^i)
\end{equation*}
\begin{equation*}
L_{RNA}=\frac{1}{m}\sum_{i=0}^{m} ||f_{gr}^i-f_{g}^i||^2
\end{equation*}
where the $f_{g}^i$ is the reconstructed feature of pathway node $i$ and the $f_{gr}$ is the initial pathway node feature. $L_{RNA}$ denotes the loss of pathway node reconstruction.\\

\noindent \textbf{Genomic knowledge-guided pathological feature extraction}
Digital pathological images contain enormous semantic information and thus have been regarded as a vital criterion of diagnosis and treatment decisions in oncology. However some normal tissues such as stroma, adipose tissue, fibroblasts exhibit less relevance to the cancer prognosis. Thus it is an intractable problem to extract prognosis-relevant information from the abundant pathological feature due to the high image heterogeneity. RNA sequencing (RNA-Seq) data serves as a direct indicator of gene expression levels and patterns within the cell nucleus, providing insights into transcriptional activity and biological function, which have been extensively linked
to the cell abnormal growth, proliferation and cancer tissue formation. Several studies have successfully applied deep learning based pathological images analysis to cancer grading \cite{kim2015knowledge}, gene mutations prediction \cite{coudray2018classification}, and gene expression prediction \cite{he2020integrating} etc, which demonstrate the feasibility of extracting gene-relevant pathological features. Thus we propose to guide the extraction of pathological feature via biological pathway information. Different from previously mentioned methods that predict the expression level or mutation status of identified key genes, we propose to conduct gene set variation analysis(GSVA) on the genomic data and supervise the pathological feature extraction with GSVA score of each pathway $f_{gsva}(m\times1)$, which can reveal the biological pathway activities in the tumor tissue. Specifically, we use attention pooling module to pathology subgraph feature embedding $F_p$ to generate the global representation of the WSI denoted as $f_{p_{wsi}}$, the global embedding $f_{p_{wsi}}$ is further input into the feedforward network to predict the GSVA score $f_{gsva}^{'}$. The GSVA prediction loss $L_{gsva}$ is calculated as follows:
\begin{equation*}
a_{i}=\frac{exp(f_0(tanh(W_0\cdot f_p^{i})\odot tanh(W_1\cdot f_p^{i})))}{\sum_{j=0}^{m}{exp(f_0(tanh(W_0\cdot f_p^{j})\odot tanh(W_1\cdot f_p^{j})))}}
\end{equation*}
\begin{equation*}
f_{p_{wsi}}=GAP(F_p)=\sum_{i=0}^{m}{a_i\cdot f_p^i}
\end{equation*}
\begin{equation*}
f_{gsva}^\prime=W_{gsva}\cdot f_{p_{wsi}}
\end{equation*}
\begin{equation*}
L_{gsva}=||f_{gsva}^{'}-f_{gsva}||^2
\end{equation*}

\noindent \textbf{Genomic and pathological feature alignment}
Pathological images and genomic data contain cancer progression-related information at distinct scales, offering diverse insights into the etiology and pathogenesis of cancer. However the heterogeneity gap between two modalities poses substantial obstacles in multi-modal fusion. Especially given that the elementary unit of genomic subgraph is biological pathways, which elaborate distinct and complicated biological process of pathological tissue. Recent work proposed to map genomic and medical image data into an unified feature space and align different modalities via contrastive learning \cite{taleb2022contig}. Based on that, we combine the feature alignment with the survival prediction to ensure that the network can be trained in an end-to-end manner with the supervision of specific downstream task. Specifically, we adopt a attention pooling module to the pathway node feature and obtain the global representation of pathway subgraph $f_{g_{pw}}$, which is then converted into the feature space of $f_{p_{wsi}}$ through linear projection and the contrastive loss between global representation of genomic subgraph $f_{g_{pw}}^\prime$ and pathology subgraph $f_{p_{wsi}}$ is as follows:
\begin{equation*}
f_{g_{pw}}^\prime=W_{pw}\cdot f_{g_{pw}}
\end{equation*}
\begin{equation*}
L_{align}=||f_{g_{pw}}^{'}-f_{p_{wsi}}||^2
\end{equation*}

\subsection {Pathology-genome heterogeneous graph learning}
Following the multi-modal representation learning module, we obtain the feature embedding of pathological nodes and pathway nodes respectively. Considering bulk RNA sequencing reveals the holistic information of genetic material within the cell nuclei of pathological tissues, each biological pathway exhibits certain associations with pathological patches, we fully connect the node pairs with different modalities in the heterogeneous graph construction. Additionally, for quantifying the intra-modality and inter-modal associations and enhancing the model interpretability, we adopt a graph learning strategy similar to graph attention network. For each node in the heterogeneous graph, we iteratively aggregate neighbor node features from intra-modality subgraph and inter-modality subgraph in each layer, and then we obtain unimodal node feature and bi-modal fused node feature from each subgraph. Take pathological node $f_p^i\ (f_p^i\in F_p)$ as an instance, we first aggregate the neighbor node $f_p^j\ \left(f_p^j\in N_p\left(i\right)\right)$ in pathology subgraph, the graph learning procedures are as follows: 1) Input the node feature $F_p$ into a linear projection layer to obtain $F_{p0}$; 2) Calculate the element-wise product of center nodes $f_{p0}^i$ with the neighbor nodes $f_{p0}^j\ (j\in N_p(i))$ and then input into linear projection layer to calculate the attention score $a_{ij}^{pp}$ of each neighbor pathological node; 3) Normalize the attention score of each neighbor nodes using the softmax function, then add the attention weighted neighbor node feature $a_{ij}^{pp}f_{p0}^j$ to the center node $f_{p0}^i$ and the updated node feature is represented as $f_{pp}^i$; 4) Input the pathway node feature $F_g$ into a linear projection layer to obtain $F_{g0}$; 5) Calculate the element-wise product of center pathological node $f_{p0}^i$ with the neighbor pathway node $f_{g0}^k\ (k\in N_g(i))$ and then input it into a linear projection layer to calculate the attention score $a_{ik}^{pg}$ of each neighbor node; 6) Normalize attention score of each neighbor nodes using the softmax function, then add the attention weighted neighbor node feature $a_{ik}^{pg}f_{g0}^k$ to the center node $f_{p0}^i$ and the updated node feature is represented as $f_{pg}^i$; The procedures of genomic node learning is similar to abovementioned steps and the formulas are as follows:
\begin{equation*}
F_{p0}=W_{p0}\cdot F_p
\end{equation*}
\begin{equation*}
a_{ij}^{pp}=\frac{exp(W_{pp}\cdot (f_{p0}^i\odot f_{p0}^j))}{\sum_{j\in N_p(i)}{exp(W_{pp}\cdot (f_{p0}^i\odot f_{p0}^j))}}
\end{equation*}
\begin{equation*}
f_{pp}^i=f_{p0}^i+\sum_{j\in N_p(i)}{a_{ij}^{pp}f_{p0}^j}
\end{equation*}
\begin{equation*}
F_{g0}=W_{g0}\cdot F_g
\end{equation*}
\begin{equation*}
a_{ik}^{pg}=\frac{exp(W_{pg}\cdot (f_{p0}^i\odot f_{g0}^k))}{\sum_{k\in N_g(i)}{exp(W_{pg}\cdot (f_{p0}^i\odot f_{g0}^k))}}
\end{equation*}
\begin{equation*}
f_{pg}^i=f_{pp}^i+\sum_{k\in N_g(i)}{a_{ik}^{pg}f_{g0}^k}
\end{equation*}

The center node can aggregate 1 hop neighbor node feature of both modalities in each graph learning step. And in our experiment the graph learning procedures are performed once which means each node in the heterogeneous graph aggregate 1 hop neighbor node from both modalities. After the above mentioned graph learning, we obtain pathological unimodal feature $F_{pp}$ and bi-modal feature $F_{pg}$ from pathological subgraph, genomic unimodal feature $F_{gg}$ and bi-modal feature $F_{gp}$ from genomic subgraph, respectively. Then we use attention pooling module to generate the global feature representations as follow:
\begin{equation*}
f_{i}=GAP(F_{i})  \quad  i\in\{pp,pg,gp,gg\}
\end{equation*}

In addition, we employ a gating-based attention mechanism \cite{arevalo2017gated} to control the weights of each features in the survival prediction task. Specifically, we use four linear projection layers of the concatenated feature $[f_{pp},f_{pg},f_{gp},f_{gg}]$ to generate four attention scores which can evaluate the relative importance of each feature embeddings. Then we integrate the gated attention weighted feature embeddings using vector concatenation followed by two fully-connected layers to predict the survival risk.
\begin{equation*}
\alpha_{i}=W_{gate}^{i}[f_{pp},f_{pg},f_{gp},f_{gg}]  \quad  i\in\{pp,pg,gp,gg\}
\end{equation*}

For each patient we denote the censorship status as c while c=0 represents the patient’s death and c=1 represents the patient had lived past the last follow-up time. We denote t as the time interval between patient’s diagnostic and the last follow-up if c=1 or the patient’s death if c=0. We denote gated attention weighted feature$\ [\alpha_{pp}f_{pp},\alpha_{pg}f_{pg},\alpha_{gg}f_{gp},\alpha_{gp}f_{gg}]$ as $s$. Besides, we adopt the negative log-likelihood (NLL) survival loss proposed by Shekoufeh \cite{zadeh2020bias} which converts the time t into equal non-overlapping time intervals $\left[t_{j-1}\ ,t_j\right],j\in\{1,2,...,q\}$, and use model to predict the label $y_j$ that represent the time interval instead of a specific time point $t_j$ and the clinical data can be represented as $[c,y_j,s]$. The log likelihood survival loss can be formulated as:
\begin{gather*}
    f_{haza}\left(y_j| s\right) = Sigmoid(\hat{y_j}) \\
    f_{surv}\left(y_j| s\right) = \prod_{k=1}^{j} (1-f_{haza}\left(y_k| s\right)) \\
    \begin{split}
    L_{surv} = & \sum_{i=0}^{N} -c^{i}\log{f_{surv}(y_j^{i}|s^{i})} \\ 
    & +(1-c^{i})\log{f_{surv}(y_j^{i}-1|s^{i})} \\
    & +(1-c^{i})\log{f_{haza}(y_{j}^{i}|s^{i})}
    \end{split}
\end{gather*}
where $\hat{y_j}$ is the predicted risk at time interval j and $N$ is the number of patients. 

The total loss is combined with survival prediction loss $L_{surv}$, genomic-pathological embedding alignment loss $L_{align}$, GSVA prediction loss $L_{gsva}$, RNA reconstruction loss $L_{RNA}$ and genomic subgraph adjacent matrix reconstruction loss $L_{Adj}$. The formula is as follows:
\begin{equation*}
L=\alpha_1L_{surv}+\alpha_2L_{align} +\alpha_3L_{gsva}+\alpha_4L_{RNA}+\alpha_5L_{Adj}
\end{equation*}
where $\alpha_i$ is the hyperparameter.

\subsection {Multimodal Interpretability}
To improve the interpretability of the model, we adopt attention visualization and gradient integration algorithms to interpret feature importance in pathological images and biological pathways, respectively. The visualization of attention weight consists of two parts: 1) attention weight obtained from pathological feature embeddings after the intra-modal learning, which aim to identify the important pathological tissue related to cancer prognosis. 2) 
pathway-pathological node co-attention generated in the heterogeneous graph learning, which reflects the correlation between pathological features and biological pathways. Besides, we also use the Integrate Gradient (IG) \cite{sundararajan2017axiomatic}, a gradient-based interpretability method, to describe how the gene expression values contribute to the prognostic risk and further identify the important biological pathways and genes with potential prognostic value. 

\section{Experiment and Results}
\subsection {Datasets and Preprocessing}
For this study, we collected whole slide images and genomic data of low grade gliomas (TCGA-LGG, n=466) and kidney renal papillary cell carcinoma (TCGA-KIRP, n=256) from The Cancer Genome Atlas (TCGA). Additionally, we gathered 181 cases of low grade gliomas (FAHZU-LGG, n=181) and 111 cases of glioblastoma (FAHZU-GBM, n=111) from First Affiliated Hospital of Zhengzhou University. Each patient contains paired pathological image and RNA expression data.\\
\indent For each cancer dataset, the pathological image preprocess procedures are as followed: we first removed the background of histology slide and divided the tissue region into non-overlapping 256×256 patches at 40× (0.5 $\mu$m/pixel), then stain normalization \cite{macenko2009method} was performed to pathological image patches to reduce the inconsistency of histology slides across different sites. Finally, we extracted feature embedding of each image patch via pretrained ResNet50. 
For genomic data, a log transformation was performed to reduce the impact of extreme values and top 20000 genes with highest variance were reserved. Then we conducted pathway enrichment with biological pathway database (KEGG and REACTOM) and preserved biological pathways with statistical significance (adj P value \textless  0.05). 480, 428 and 472 pathways were obtained in LGG, GBM and KIRP dataset respectively. The gene set variation analysis was then conducted and the GSVA scores of biological pathways were calculated to guide the pathological feature extraction.
The models in our experiments were trained using 5-fold cross validation. The TCGA-KIRP, FAHZU-GBM and TCGA-LGG dataset were used for model training and validation while the FAHZU-LGG dataset served as an external validation dataset to evaluate the generalization of models trained in TCGA-LGG dataset.

\subsection {Implementation Details}
\noindent \textbf{Parameter Setting}
The pathological features were extracted via ResNet50 with the pretrained weights from ImageNet \cite{he2016deep} and the 1024-dimensional feature embeddings were obtained from the average pooling layer behind the 3rd residual block. Considering the biological pathways contained different number of RNA, we first adopted encoder networks to each pathway to obtain the unified dimensional pathway features. Then feedforward network which consisted of two fully connected layers, ReLU activation and Dropout layers (p=0.2) was adopted to pathology feature embeddings to unify the feature dimension of both modalities. In heterogeneous graph learning stage, the node feature of pathological image patch and genomic pathway aggregated neighbor node feature from intra-modal subgraph and inter-modal subgraph consecutively. Considering the edges of inter-modality were built through fully connecting, each node feature only aggregated 1-hop neighbor nodes from each modality. The following ablation experiment and comparison experiment were conducted in the same train/valid splits. All networks were trained using Adam optimizer with weight decaying of $1\times{10}^{-4}$ and learning rate of $2\times{10}^{-4}$. The batch size is 1 with 8 gradient accumulation steps.

\noindent \textbf{Model Evaluation}
We evaluated our proposed method with standard quantitative and statistical metrics for survival analysis. Specifically, we evaluated the prediction performance using the Concordance Index (CI) which represents the proportion of patient pair that predicted risk is ranked in the same order as survival time among all uncensored patients. We calculated the CI value of each model trained in 5-fold cross validation and the model performance was given in terms of mean and standard deviation. Besides, we also conducted the Kaplan–Meier (KM) analysis to visualize the patient stratification result. For assessing the significance of stratification, the log rank test \cite{bland2004logrank} was utilized to determine the statistical significance of the difference between low and high risk patient group.

\subsection {Ablation Study}
We conducted an ablation study to evaluate the performance of unimodal networks including pathological subgraph and genomic subgraph to verify the effectiveness of multi-modal fusion. Besides, to demonstrate the performance improvement of biological knowledge guided representation learning module, we also trained the network without the module using the same parameter settings. And Table  \uppercase\expandafter{\romannumeral1} has shown the result of ablation experiments that the multimodal fusion models yielded better performance comparing to both unimodal subgraph in each cancer dataset. Meanwhile the PathoGenoSurvGraph with biological knowledge guided representation learning module achieved the consistently highest c-Index. For the low-grade gliomas dataset, the performance of genomic subgraph surpass the pathological subgraph and the same result is observed in the TCGA-KIRP dataset. And the PathoGenoSurvGraph achieves 2.1$\%$ and 4.9$\%$ performance increase comparing to the best unimodal model in TCGA-LGG and FAHZU-LGG dataset, respectively. For FAHZU-GBM dataset, we observe that the pathological subgraph achieves better performance than the genomic subgraph, which is different from the other datasets and PathoGenoSurvGraph also achieves 4.5$\%$ performance gain comparing with the pathological subgraph. Furthermore, we also observe 4.9$\%$ and 3.5$\%$ performance improvement comparing with the best performance of unimodal model and the PathoGenoSurvGraph without the biological knowledge guided representation learning module in TCGA-KIRP dataset.

\begin{table}[!htbp] 
    \renewcommand\arraystretch{1.2}
    \captionsetup{justification=centering} 
    \caption{Concordance index of PathoGenoSurvGraph and ablation experiments in cancer survival prediction}
    \fontsize{8}{12}\selectfont 
    \setlength{\tabcolsep}{1pt} 
    \centering 
    \begin{threeparttable}          
    \begin{tabular}{>{\centering}p{1.85cm}|>{\centering}p{1.7cm}>{\centering}p{1.8cm}|>{\centering}p{1.8cm}|>{\centering}p{1.7cm}} 
        \Xhline{1px}
        Model & TCGA-LGG & FAHZU-LGG$^{*}$ & FAHZU-GBM  & TCGA-GBM \\ 
        \hline  
        PathoGraph & 0.602$\pm$0.037 & 0.534$\pm$0.021 & 0.574$\pm$0.034 & 0.661$\pm$0.087 \\  
        GenoGraph  & 0.806$\pm$0.021 & 0.653$\pm$0.035 & 0.531$\pm$0.038 & 0.737$\pm$0.079 \\
        \hline  
        PGHG (wo BG) & 0.806$\pm$0.021 &0.655$\pm$0.025 &0.596$\pm$0.030 & 0.747$\pm$0.114 \\  
        PGHG (w BG) & \textbf{0.823$\pm$0.026} & \textbf{0.685$\pm$0.011}&  \textbf{0.600$\pm$0.032} 
        &  \textbf{0.773$\pm$0.109} \\  
        \Xhline{1px}
    \end{tabular}
    \begin{tablenotes}    
    \item[] w/wo BG: with/without the biological knowledge guided representation learning module. 
    \end{tablenotes} 
    \end{threeparttable}          
\end{table}
Through the ablation experiment, we demonstrate that the integration of pathological image feature and the genomic feature can consistently achieve performance improvement in each cancer dataset. And the additive performance gain of adopting the biological knowledge guided representation learning module suggests that biological knowledge is informative to guide the extraction of semantically
meaningful information related to cancer prognosis. We also conduct the Kaplan-Meier (KM) analysis to visualize the performance of patient stratification, and then log rank test is used to test the statistical significance of high-low risk group. Figure 1 shows that the KM curves of pathological subgraph, genomic subgraph, PathoGenoSurvGraph w/wo biological knowledge guided representation learning module in each dataset.

\renewcommand\arraystretch{1.5} 
\begin{figure*}[tp] 
\centering 
\includegraphics[width=5.5in]{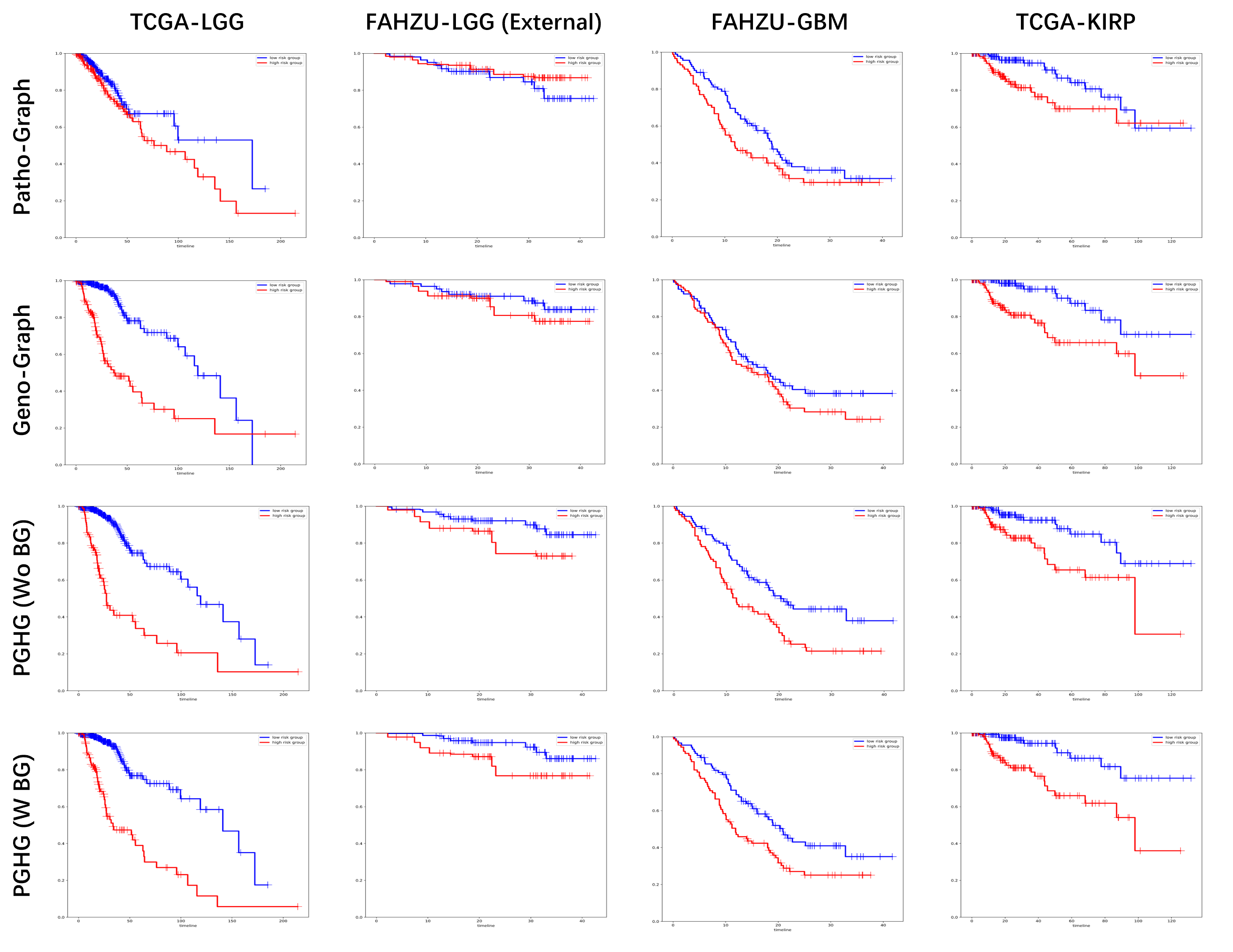}
\captionsetup{justification=centering} 
\caption{Kaplan Meier curves of PathoGenoSurvGraph, includes pathological subgraph, genomic subgraph and PathoGenoSurvGraph w/wo biological knowledge guided representation learning module.}
\label{amplifier}
\end{figure*}

\begin{table*}[bp] 
    \renewcommand\arraystretch{1.2}
    \centering  
    \captionsetup{justification=centering} 
    \caption{Concordance index of PathoGenoSurvGraph and comparison experiments in cancer survival prediction}
    \fontsize{8}{12}\selectfont 
    \setlength{\tabcolsep}{1pt} 

    \centering 
    \begin{tabular}{>{\centering}p{2cm}|>{\centering}p{1cm}|>{\centering}p{1cm}|>{\centering}p{1.7cm}>{\centering}p{1.8cm}|>{\centering}p{1.8cm}|>{\centering}p{1.7cm}} 
        \Xhline{1px}
        Model & Patho & Geno & TCGA-LGG & FAHZU-LGG$^*$ & FAHZU-GBM  & TCGA-GBM \\ 
        \hline  
        MLP & \usym{2717} & \usym{1F5F8} & 0.799$\pm$0.040 & 0.597$\pm$0.015 & 0.556$\pm$0.096 & 0.625$\pm$0.112 \\
        SNN & \usym{2717} & \usym{1F5F8} & 0.809$\pm$0.035 & 0.641$\pm$0.025 & 0.550$\pm$0.073 & 0.637$\pm$0.123 \\
        \hline  
        AttenMIL & \usym{1F5F8} & \usym{2717} & 0.591$\pm$0.066 & 0.643$\pm$0.009 & 0.511$\pm$0.044 & 0.592$\pm$0.082 \\
        TransMIL & \usym{1F5F8} & \usym{2717} & 0.600$\pm$0.031 & 0.534$\pm$0.025 & 0.552$\pm$0.038 & 0.652$\pm$0.085 \\
        \hline  
        AttenMIL+SNN & \usym{1F5F8} & \usym{1F5F8} & 0.822$\pm$0.022  & 0.673$\pm$0.012  & 0.526$\pm$0.074  & 0.743$\pm$0.128 \\
        TransMIL+SNN & \usym{1F5F8} & \usym{1F5F8} & 0.819$\pm$0.017  & 0.680$\pm$0.011  & 0.545$\pm$0.064  & 0.761$\pm$0.110 \\
        SurvPath & \usym{1F5F8} & \usym{1F5F8} & 0.798$\pm$0.024 &  0.654$\pm$0.016  & 0.578$\pm$0.035  & 0.661$\pm$0.107 \\
        \hline  
        PGHG (wo BG) & \usym{1F5F8} & \usym{1F5F8} & 0.806$\pm$0.021&  0.655$\pm$0.025&  0.596$\pm$0.030&  0.747$\pm$0.114 \\
        PGHG (w BG) & \usym{1F5F8} & \usym{1F5F8} & \textbf{0.823$\pm$0.026} & \textbf{0.685$\pm$0.011}&  \textbf{0.600$\pm$0.032}&  \textbf{0.773$\pm$0.109} \\  
        \Xhline{1px}
    \end{tabular}
\end{table*}

The result shows that PathoGenoSurvGraph w/wo biological knowledge guided representation learning module consistently achieve superior stratification performance compared to the pathological subgraph and genomic subgraph. And the KM curves of external dataset FAHZU-LGG, which has relatively shorter follow-up time and thus higher stratification difficulty, also shows statistically significance, thus further reveals the effectiveness and generalization of PathoGenoSurvGraph.

\subsection {Comparison Experiment}
In comparison with other state of the art models, we conducted comparison experiment with other unimodal and multimodal strategies in each datasets. In unimodal model comparison, we adopted Multi-Layer Perceptron (MLP) and Self-Normalizing Network (SNN) \cite{klambauer2017self} to genomic data which has extremely high feature dimension. The SNN consisted of linear projection layer, SeLU activation and Alpha Dropout. And for pathological unimodal model comparison, we adopted the AttenMIL \cite{ilse2018attention} with gated-attention pooling and TransMIL \cite{shao2021transmil} with Nystrom attention pooling \cite{xiong2021nystromformer}. In the multi-modal comparison experiment, late fusion strategy was used to combine AttenMIL and TransMIL with SNN through vector concatenation to fuse pathological and genomic feature, besides we also compared with the transformer-based method SurvPath which utilized the cross-attention matrix of pathological image and genomic pathway to guide the fusion of both modalities. In the comparison experiment, all models used the same hyperparameters and five-fold cross-validation. In total, we trained 220 models in the ablation experiment and the comparison experiment and the result is showed in the Table \uppercase\expandafter{\romannumeral2}.

\renewcommand\arraystretch{1.5} 
\begin{figure*}[tp] 
\centering 
\includegraphics[width=5.5in]{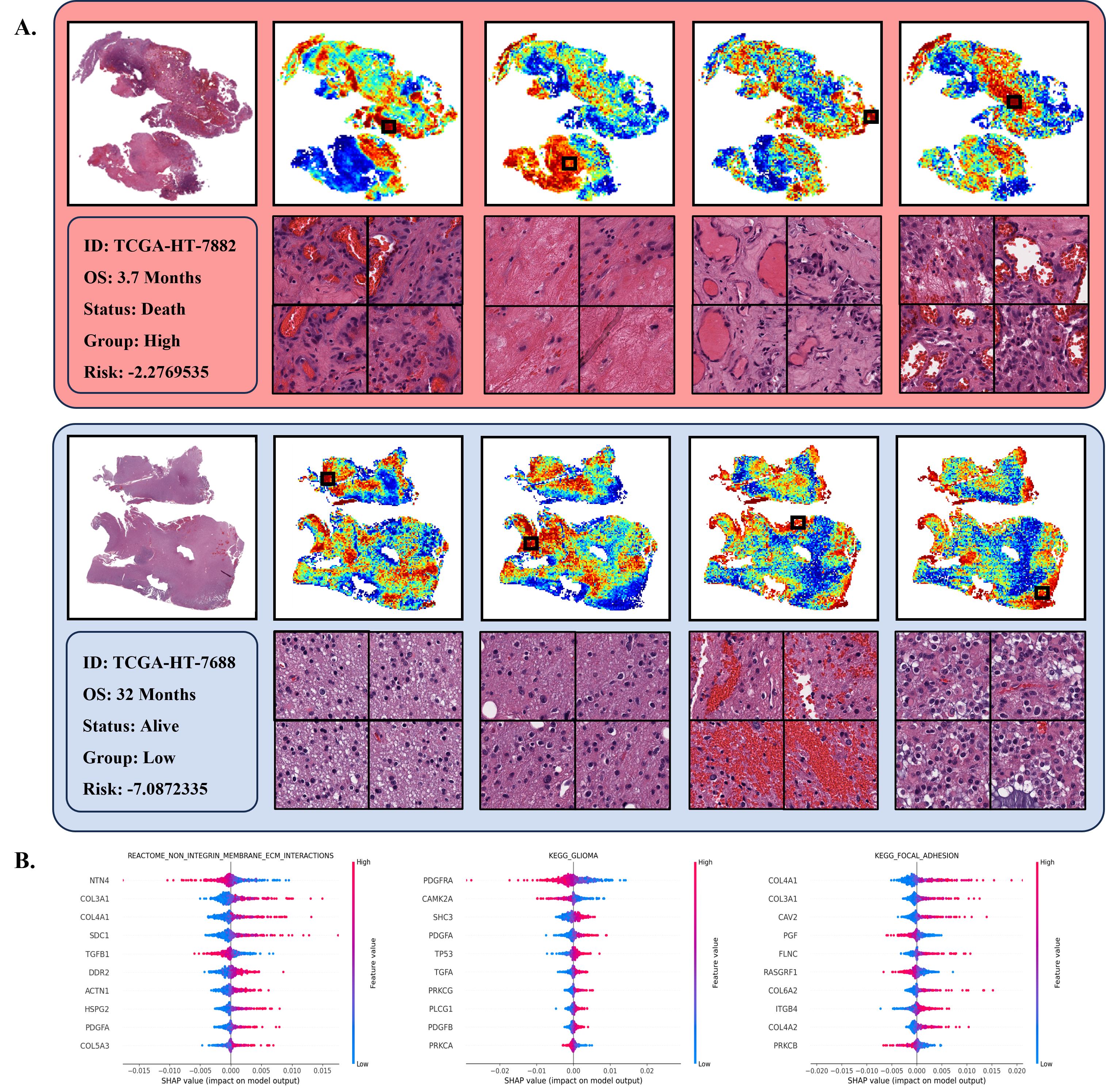}
\caption{Genomic and histological interpretability in low grade gliomas. A: Global attention visualization and three high mean absolute IG biological pathways (REACTOME NON INTEGRIN MEMBRANE ECM INTERACTIONS, KEGG GLIOMA and KEGG FOCAL ADHESION) co-attention visualization for high-risk patient and low-risk patient in TCGA-LGG dataset. B: Top 10 absolute IG value RNA in three biological pathways with the color indicates the relative expression color.}
\label{amplifier}
\end{figure*}

The result shows that our method achieve superior performance than all other unimodal and multimodal models. In the TCGA-LGG dataset, we observed that the genomic unimodal models have better performance than the pathological unimodal models while the opposite was observed in the FAHZU-GBM dataset, which was in accordance with the result in ablation experiment. The consistently performance improvement achieved by the fusion of pathological and genomic features also demonstrated the effectiveness of our multimodal fusion strategy in survival prediction task. For the TCGA-LGG dataset and the external FAHZU-LGG dataset, the PathoGenoSurvGraph outperformed the best unimodal model with 1.7$\%$ and 6.5$\%$ improvements respectively, reaching c-Index of 0.823 and 0.685. Simultaneously, compared to other multimodal models based on late fusion strategy, it also achieved a modest performance improvement. Similarly, in the TCGA-KIRP, all of the multimodal models achieved superior performance compared with the unimodal model. And comparing to the best multi-modal network, the PathoGenoSurvGraph also achieved the performance increase of 1.6$\%$.

\renewcommand\arraystretch{1.5} 
\begin{figure*}[tp] 
\centering 
\includegraphics[width=5.5in]{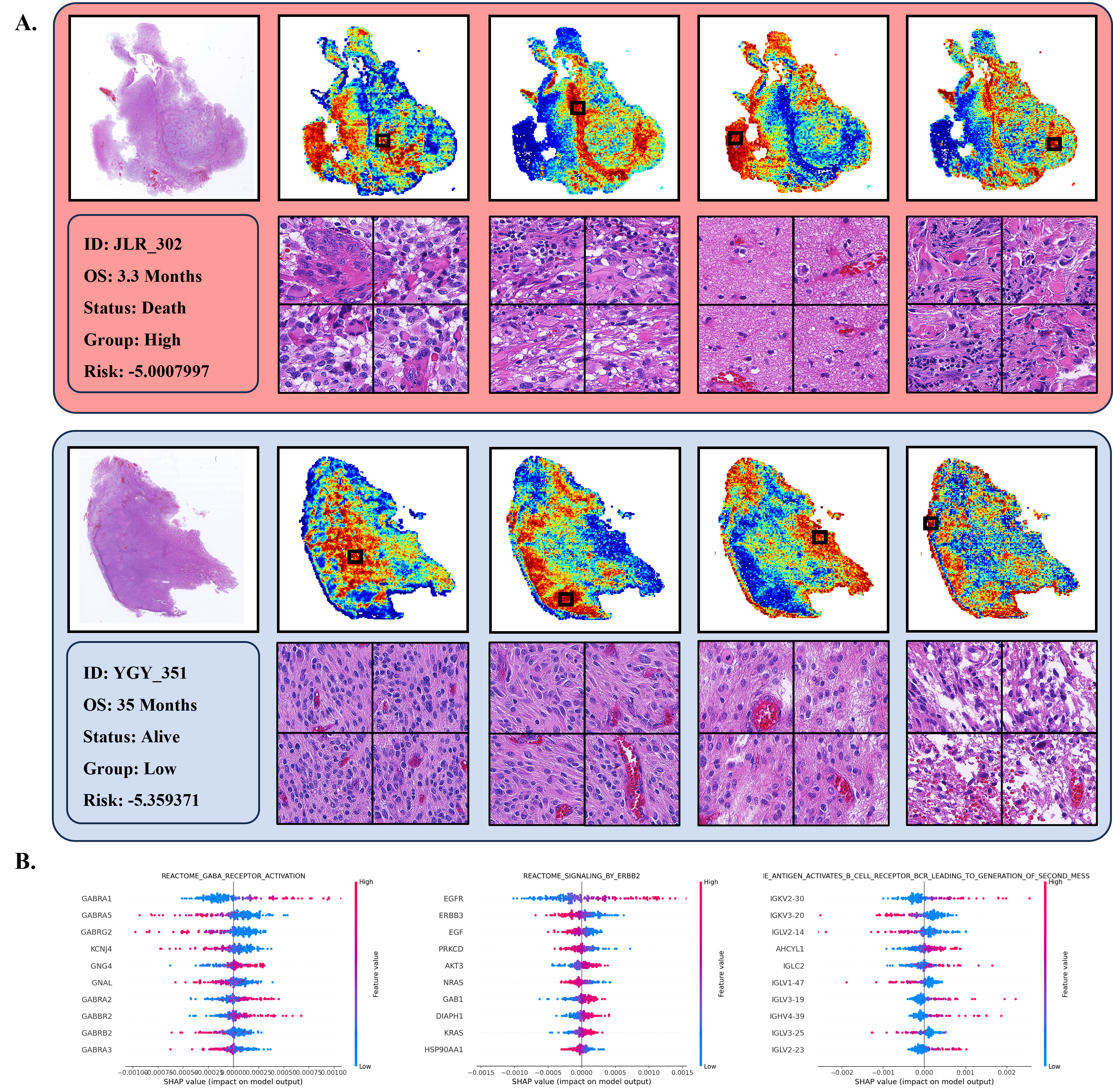}
\caption{Genomic and histological interpretability in glioblastoma. A: Global attention visualization and three high mean absolute IG biological pathways (REACTOME SIGNALING BY ERBB2, REACTOME ANTIGEN ACTIVATES B CELL RECEPTOR BCR LEADING TO GENERATION OF SECOND MESSENGERS and REACTOME GABA RECEPTOR ACTIVATION) co-attention visualization for high-risk patient and low-risk patient in FAHZU-GBM dataset. B: Top 10 RNA in each biological pathways with the color indicates the relative expression value.}
\label{amplifier}
\end{figure*}

\subsection {Multimodal Interpretability}
We also demonstrated the interpretability of our method. Firstly, we obtained the attention value from the attention pooling layer and generated the attention heatmap of WSIs from high and low risk group to identify the morphological features with prognostic value. To explore the potential genomic biomarkers relevant to cancer prognosis, we selected three key pathways from the top 25 pathway with the highest IG value. Specifically, we visualized the co-attention heatmap to demonstrate the association with tumor tissue and identified the key genes through the IG values. The Figure below contains the WSI global attention heatmap, key pathway co-attention heatmap and Top 10 genomic features with the highest IG value.

\renewcommand\arraystretch{1.5} 
\begin{figure*}[tp] 
\centering 
\includegraphics[width=5.5in]{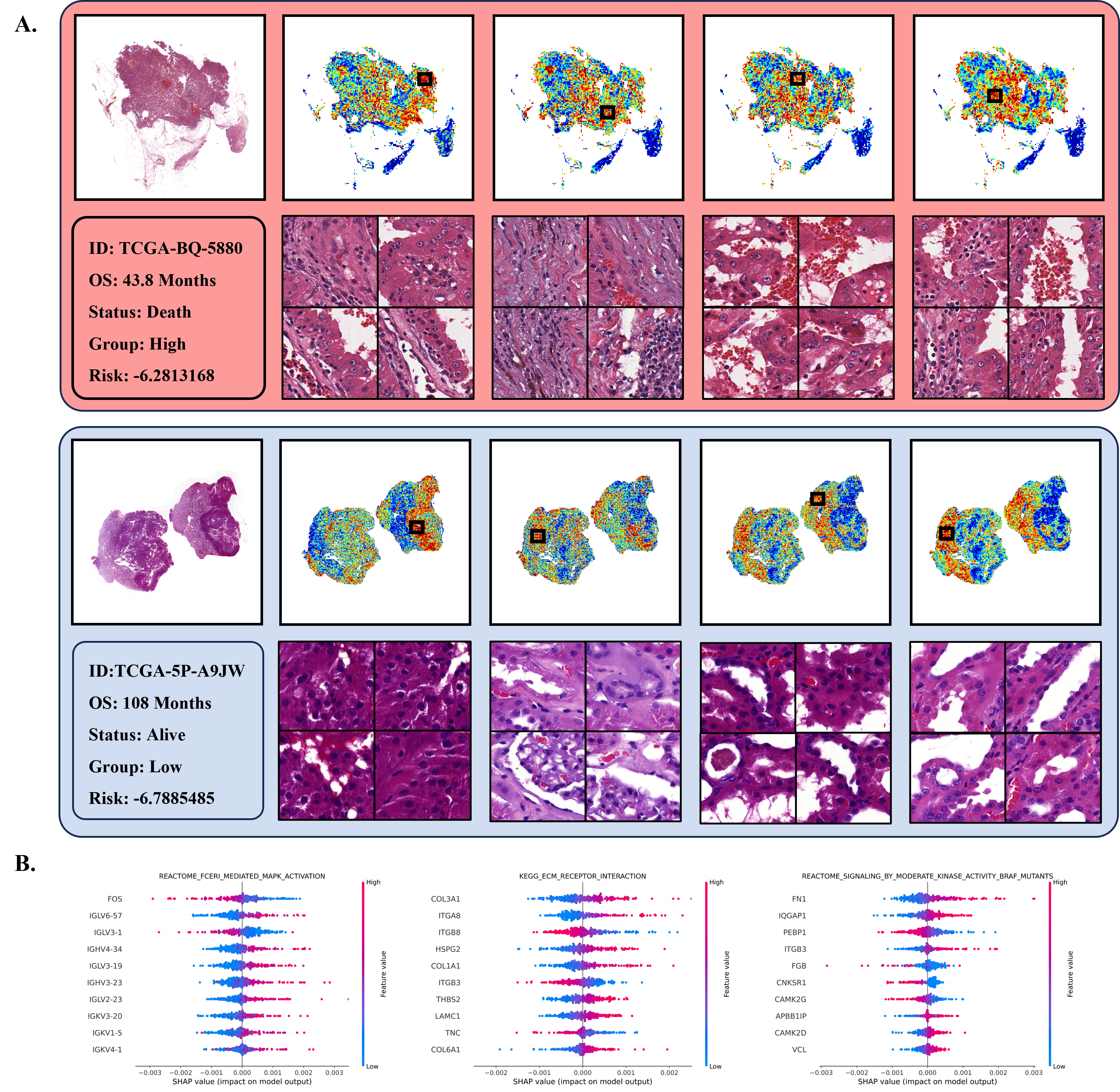}
\caption{Genomic and histological interpretability in kidney renal papillary cell carcinoma. A: Global attention visualization and three high mean absolute IG biological pathways (REACTOME FCERI MEDIATED MAPK ACTIVATION, KEGG ECM RECEPTOR INTERACTION and REACTOME SIGNALING BY MODERATE KINASE ACTIVITY BRAF MUTANTS) co-attention visualization for high-risk patient and low-risk patient in TCGA-KIRP dataset. B: Top 10 RNA in each biological pathways with the color indicates the relative expression value.}
\label{amplifier}
\end{figure*}

\noindent \textbf{TCGA-LGG}
The global attention heatmap of both high risk patient and low risk patient indicates that the high-density gliomas cell and microvascular proliferation are more informative for the survival prediction. For the key pathway co-attention heatmap, we observe that \seqsplit{REACTOME\_NON\_INTEGRIN\_MEMBRANE\_ECM\_INTERACTIONS} pathway has relatively high mean absolute IG values, which is known to relate to the interaction of glioma cells with the extracellular matrix and the subsequent destruction of matrix barriers, the shift of gene expression in this pathway may further initiate or influence the process of glioma cell invasion \cite{goldbrunner1999cell}, the corresponding co-attention heatmap demonstrates the association of the pathway with high-density tumor cells and necrosis area. Besides, our approach also highlights the several prognostic markers such as NTN4, PDGFA. Specifically, the overexpression of NTN4 can act as an anti-inflammation factor in endothelial cell and decrease the risk \cite{ozhan2021smultcan}, which we similarly observe that higher NTN4 expression can improve prognosis. Besides, we also observe that increasing the expression of COL4A1 and PDGFA can increase the risk, which is consistent with previous study \cite{vastrad2017molecular}. KEGG\_FOCAL\_ADHESION also involves with the interaction with downstream targets of integrins and growth factor receptors, thus further influence the survival, proliferation and invasion of tumor cell, the corresponding co-attention heatmap demonstrates that the pathway is closely related to regions with tumor cell. FLNC, COL4A1 and RASGRF1 are also highly attributed that increase FLNC \cite{kamil2019high}, COL4A1 \cite{vastrad2017molecular} expression and decrease RASGRF1 expression can increase the risk of cancer death \cite{li2021molecular}.

\noindent \textbf{FAHZU-GBM}
In FAHZU-GBM dataset, the global attention heatmap also allocates large attention weight on the regions with tumor cells. Besides, the majority of important pathways relates to cell invasion, apoptosis, and cancer prognosis. For instance, REACTOME\_SIGNALING\_BY\_ERBB2 pathway has been implicated in regulating cell proliferation, migration and apoptosis and the corresponding high attention histological tissue exhibiting the presence of high-density gliomas cells and microvascular, which can provide efficient blood supply of tumor growth and metastasis \cite{yu2011mir}. From the top 10 IG gene plot, we can observe numerous oncogenes and prognostic markers such as EGFR \cite{cancer2015comprehensive}, GAB1 \cite{rosell2008translational} which have significant impact on the tumor progression, the overexpression of both genes are significantly associated with progression, proliferation, and metastasis across many cancers \cite{jafarzadeh2022microrna}. Besides the expression level of GAB1 also correlates with cellular proliferation, evasion of apoptosis and angiogenesis. REACTOME\_GABA\_RECEPTOR\_ACTIVATION can regulate the growth of malignant tumors with the corresponding co-attention heatmap concentrating on the high-density tumor cell and necrosis area. Besides, the model discovers that increasing GNG4 \cite{jiang2023high} expression can lead to poor prognosis, which is also in line with existing literature.

\noindent \textbf{TCGA-KIRP}
The global attention heatmap mainly concentrates on the fibrovascular cores on high risk patient while also attends to large areas of hemorrhage in low risk group. Besides, the model allocates high attribute value to the gene in REACTOME\_FCERI\_MEDIATED\_MAPK\_ACTIVATION pathway which is critical to regulate the biological response of immune cells, the co-attention heatmap attends towards tall columnar tumor cells and fibrovascular cores area in both cases. The high IG gene plot also demonstrates that key biomarker such as FOS which has shown downregulation across many cancers \cite{li2022machine,kondapuram2022pan}, can greatly influence the cancer prognosis. Besides, \seqsplit{REACTOME\_SIGNALING\_BY\_MODERATE\_KINASE\_ACTIVITY\_BRAF\_MUTANTS} and \seqsplit{KEGG\_ECM\_RECEPTOR\_INTERACTION} pathways are also identified to be significant in predicting kidney renal papillary cell carcinoma prognosis, while the former pathway has been demonstrated that can influence cancer progression through regulating certain kinase activity and the latter pathway involves in the regulation of cell migration and invasion.\\
\indent The above-mentioned result demonstrates that our model can uncover the pivotal biological pathway and genes which may serve as the biomarkers in cancer diagnosis and prognosis. Besides, the pathway-histology co-attention analysis also reveal the association between both modalities, which can be used in cancer pathogenesis study.

\section{Conclusion}
In this work, we propose a heterogeneous graph framework to establish the correlation between histological image and genomic pathway. We represent the histological patches and genomic pathways as nodes in each subgraph and construct the edges based on the spatial location of patches and the common gene number of genomic pathways respectively. The edges between each subgraph are fully connected to identify the correlation of pathological tissue and genomic data. After the graph construction, we design the biological prior knowledge-based loss function for the genomic and pathological feature extraction and adopt graph attention mechanism for heterogeneous graph training. Then we use the global attention pooling module to extract the unimodal and bi-modal feature embeddings from each subgraph and make the survival prediction. The model is evaluated in LGG and KIRP dataset from TCGA, GBM dataset from FAHZU and an external validation dataset of LGG from FAHZU and achieves promising performance in comparisons with other methods. Besides, the heterogeneous graph network is scalable and interpretable which can incorporate with diverse modalities of clinical data and uncover the potential and novel biomarkers.

\ifCLASSOPTIONcaptionsoff
  \newpage
\fi






\bibliographystyle{IEEEtran}
\bibliography{IEEEabrv,Bibliography}






\vfill


\end{sloppypar}

\end{document}